\documentclass[%
 reprint,
superscriptaddress,
 amsmath,amssymb,
 aps,
 prx,
]{revtex4-2}
\usepackage{graphicx}
\usepackage{dcolumn}
\usepackage{bm}
\usepackage{physics}
\usepackage{hyperref}
\usepackage[caption=false]{subfig}
\usepackage{booktabs}
\usepackage{moreverb}
\usepackage{color}

\newcommand{\mE}{\mathcal{E}}

\immediate\write18{texcount -inc -incbib 
-sum borra.tex > /tmp/wordcount.tex}

\newcommand{\mI}{{\bm I}}

\newcommand{\mL}{{\bm L}}

\makeatletter
\newcommand*{\addFileDependency}[1]{
  \typeout{(#1)}
  \@addtofilelist{#1}
  \IfFileExists{#1}{}{\typeout{No file #1.}}
}
\makeatother

\immediate\write18{texcount -char -freq
 borra.tex > /tmp/charcount.tex}

\usepackage{amsthm}

\theoremstyle{plain}

\newtheorem*{theorem*}{Theorem}

\usepackage{float}
\usepackage[ruled,linesnumbered]{algorithm2e}

\begin{document}

\preprint{arXiv}

\title{Learning the eigenstructure of quantum dynamics using classical shadows}


\author{Atithi Acharya}
\affiliation{Department of Physics and Astronomy, Rutgers University, Piscataway, NJ 08854, USA}
\author{Siddhartha Saha}
\affiliation{Department of Physics and Astronomy, Rutgers University, Piscataway, NJ 08854, USA}
\author{Shagesh Sridharan}
\affiliation{Department of Physics and Astronomy, Rutgers University, Piscataway, NJ 08854, USA}
\author{Yanis Bahroun}
\affiliation{
Center for Computational Neuroscience, Flatiron Institute New York, NY 10010 USA}
\affiliation{Center for Computational Mathematics,Flatiron Institute New York, NY 10010 USA}  
\author{Anirvan M. Sengupta}
\affiliation{Department of Physics and Astronomy, Rutgers University, Piscataway, NJ 08854, USA}
\affiliation{Center for Computational Mathematics,Flatiron Institute New York, NY 10010 USA}  \affiliation{Center for Computational Quantum Physics, Flatiron Institute, New York, NY 10010 USA}

\begin{abstract}
Learning dynamics from repeated observation of the time evolution of an open quantum system, namely, the problem of quantum process tomography is an important task. This task is difficult in general, but, with some additional constraints could be tractable. This motivates us to look at the problem of Lindblad operator discovery from observations. We point out that for moderate size Hilbert spaces, low Kraus rank of the channel and short time steps, the eigenvalues of the Choi matrix corresponding to the channel have a special structure. We use the least-square method for the estimation of a channel where, for fixed inputs, we estimate the outputs by classical shadows. The resultant noisy estimate of the channel can then be denoised by diagonalizing the nominal Choi matrix, truncating some eigenvalues, and altering it to a genuine Choi matrix. This processed Choi matrix is then compared to the original one. We see that as the number of samples increases, our reconstruction becomes more accurate. We also use tools from random matrix theory to understand the effect of estimation noise in the eigenspectrum of the estimated Choi matrix.

\end{abstract}

\maketitle

\section{Introduction}

Learning dynamics from observations is an important task in many fields. For open quantum systems, this problem is called Quantum Process Tomography (QPT).
In standard QPT, the process is an unknown Completely Positive Trace-Preserving (CPTP) map on operators associated with a $d$-dimensional Hilbert space. The CPTP map requires $d^4-d^2$ real numbers to be completely characterized \cite{Nielsen}. For $n$-qubit sytems, $d=2^n$, so the number of parameters is $O(2^{4n})$. Thus, even for a 10-qubit system, QPT formally requires estimating about $10^{12}$ parameters well, necessitating a large number of observations. However, if we have prior information that the map is close to identity, with the nontrivial action due to a small number of Lindblad operators \cite{Nielsen}, we might make some progress. This paper explores the conditions under which such progress is possible.

For quantum state tomography (QST), shadow tomography \cite{aaronson2018shadow} aims at predicting a power law number of observations in the number of qubits, $n$, from $O(n)$ copies of the density matrix $\rho$. The authors in \cite{Huang_2020} have constructed such a description of low sample complexity via the so-called classical shadows. \cite{acharya2021shadow} extended this method to generalized measurements.

Recently, the work done in \cite{levy2021classical} and \cite{kunjummen2021shadow} used Choi-Jamio\l{}kowski correspondence between channels and states to apply the classical shadows technique to QPT. However, since classical shadows do not produce the state, and therefore in the QPT context do not give you the channel, it is not clear how to perform general dynamical prediction over longer time scales. This motivates us to look at the problem of Lindblad generator discovery \cite{boulant2003robust, Howard_2006} while using classical shadow tomography for state estimation.

A loosely related subject in machine learning is the Lie generator \cite{rao1998learning, miao2007learning}, which has seen a resurgence of interest \cite{hashimoto2017unsupervised, bahroun2019similarity}.
In subsequent studies, such as \cite{hashimoto2017unsupervised, bahroun2019similarity}, the spectral gap is crucial to identify the number of truly active generators. We also observe a comparable significance of the spectral gap in our work.

Recently, least square methods have been used for QST with classical shadows \cite{nguyen2022optimising}.
Our approach utilizes least square method for the estimation of a channel where, for fixed inputs, we estimate the output by classical shadows. The resultant noisy estimate of the channel can then be diagonalized/factorized. Under appropriate circumstances, a spectral gap allows us to truncate the factorized version and essentially denoise the estimate.


\section{Quantum channel as a linear map and its factorization}
\label{sec:QC_linear}


Let the quantum channel be defined as a map $\mE:\mathbb{C}^{d\times d}\to \mathbb{C}^{d\times d}$ with the input and output density matrix related by $ \rho^{out}=\mE(\rho^{in})$. Explicitly, in terms of components, we have:
\begin{equation}
    \rho^{out}_{ij}=\sum_{kl}\mE_{ijkl}\rho^{in}_{kl}.
\end{equation}
Jamio\l{}kowski-Choi correspondence \cite{jamiolkowski1972linear,choi1975completely} creates a density matrix,  the Choi matrix,  on $\mathbb{C}^{d}\otimes \mathbb{C}^{d}$:
\begin{equation}
    \mathbf{C}_\mE=\sum_{ij}\ket{i}\bra{j}\otimes \mE(\ket{i}\bra{j}).
\end{equation}
Note that the Choi matrix elements are $$(\mathbf{C}_\mE)_{(ik),(jl)}=\bra{k}\bra{i}\mathbf{C}_\mE\ket{l}\ket{j}=\bra{k}\mE(\ket{i}\bra{j})\ket{l}=\mE_{klij}.$$

The Choi matrix is positive semidefinite and has an eigendecomposition of the form $\mathbf{C}_\mE=\sum_\alpha\lambda_\alpha \mu_\alpha \mu_\alpha^\dagger$ 
with $\mu_\alpha\in \mathbb{C}^{d}\otimes \mathbb{C}^{d}$ and $\lambda_\alpha\ge 0$ for all $\alpha$. Defining $M^\alpha_{ik}=\sqrt{\lambda_\alpha}(\mu_\alpha)_{ki}$, we get  a factorization of our the channel \cite{Howard_2006} $\mE$ :    
\begin{equation}
\mE_{ijkl}=\sum_{\alpha}M^{\alpha}_{ik}(M^{\alpha}_{jl})^*. \end{equation}
The number of nonzero eigenvalues of the Choi matrix is the Kraus rank of the channel. This factorization is the basis of the operator-sum representation \cite{Nielsen}: $\rho^{out}=\sum_\alpha M^\alpha\rho^{in}M^{\alpha\dagger}$. For systems with low Kraus rank, we can use this factorization to denoise estimated channels.

\def\mE{\mathcal {E}}
\def\mL{\mathcal{L}}


We proceed by defining a straightforward loss function:
\begin{equation}
\mathcal{L}=\sum^{T}_{t=1}\left( \rho^{out}_{ij}(t)-\sum_{kl}\mE_{ijkl}\rho^{in}_{kl}(t) \right)^{2}
\end{equation}
where $t$ serves as a sample index. By setting the derivative, $\frac{\partial \mathcal{L}}{\partial \mE_{ijcd}}$, to zero, we derive the optimal channel estimate:
\begin{equation}
    \sum^{T}_{t=1}\rho^{out}_{ij}(t)\rho^{in}_{cd}(t)=\sum_{kl}\hat\mE_{ijkl}\sum^{T}_{t=1}\rho^{in}_{kl}(t)\rho^{in}_{cd}(t)
\end{equation}
This representation elucidates the terms as non-centered covariance expressions. For example, we define the first covariance expression that captures the overlap between the input and output states, i.e. $C^{out,in}$, and the second covariance expression that captures the overlap between the different input states, i.e. $C^{in,in}$ as 

\begin{subequations}
\begin{align}
     C^{out,in}_{ij, cd}=\frac{1}{T}\sum^{T}_{t=1}\rho^{out}_{ij}(t)\rho^{in}_{cd}(t) \\
     C^{in,in}_{kl, cd} = \frac{1}{T} \sum^{T}_{t=1} \rho^{in}_{kl}(t)\rho^{in}_{cd}(t) 
\end{align} 
\end{subequations}

We can further expand the first term in terms of the second, $C^{out,in}_{ij, cd}=\sum_{kl}\epsilon_{ij, kl}C^{in, in}_{kl, cd}$. 

To simplify the notation, we group the indices as follows: $ij \rightarrow \alpha$, $cd\rightarrow \beta$, and $kl \rightarrow \gamma$. This results in:
\begin{equation}
    C^{out,in}_{\alpha, \beta}=\sum_{\gamma}\hat\mE_{\alpha \gamma}C^{in, in}_{\gamma \beta}
\end{equation}
Interpreting this in the form of a matrix equation aids in determining the quantum channel estimate: $\mathbf{C}^{out,in}=\hat\mE\mathbf{C}^{in,in}$. Under the assumption that $\mathbf{C}^{in,in}$ is invertible, we arrive at the following factorization:
\begin{equation}
    \hat\mE=\mathbf{C}^{out,in}(\mathbf{C}^{in,in})^{-1}.
    \label{Eq:E_C}  
\end{equation}

\begin{figure}
    \centering
    \includegraphics[width=8cm]{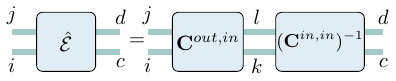}
    \caption{Expressing the quantum channel \eqref{Eq:E_C} in terms of the covariance expressions $C^{out, in}$ and inverse of  $C^{in, in}$. This factorization is explained in detail in the Sec.\ref{sec:QC_linear}. }
    
    \label{fig:E_eqn}
\end{figure}


\section{Eigenstructure of low rank channels}
For open quantum systems, the dynamics of a quantum system is described by Lindblad master equation \cite{gorini1976completely,lindblad1976generators},

\begin{gather}
\label{Eq:master_eqn}
\nonumber
\dot{\rho}(t)=-i[\mathcal{H}, \rho(t)]+\sum_{k=1}^{N}\biggl( L_{k} \rho(t) L_{k}^{\dagger}-\frac{1}{2} L_{k}^{\dagger} L_{k} \rho(t)-  \\ \frac{1}{2} \rho(t) L_{k}^{\dagger} L_{k} \biggl)
\end{gather}

\def\dt{\Delta t}

We also summarize further details in \ref{sec:Appendix_GKSL}.
We note that for the channel $ \rho^{out} = \mE(\rho^{in}) = \sum_{\alpha} M_{\alpha} \rho^{in} M_{\alpha}^{\dagger}$, 
we have the following:
\begin{equation}
    \mE_{ijkl}=\sum_{\alpha}M^{\alpha}_{ik}M^{\alpha^*}_{jl} 
\end{equation}

To relate this to the discrete-time channel description, we note that the appropriate factorized structure in the four-index tensor $\mE_{ijkl}$ is manifested when the indices $(i,k)$ and $(j,l)$ are clubbed together. Eigendecomposition of ${\mathbf{C}_\mE}_{(i,k),(j,l)}$ gives us the Kraus operators $M_{\alpha}$ and the Kraus rank denotes the number of nonzero eigenvalues of the Choi matrix. For a discrete-time version of the Lindblad master equation we consider the following channel:
\begin{equation}
     \rho^{out} = \mE(\rho^{in}) = (1-p)UM\rho^{in} M^{\dagger}U^{\dagger} + p\sum_{\alpha}L_{\alpha}\rho^{in} L_{\alpha}^{\dagger}
\end{equation}
where
\begin{equation}
     M = \left(\frac{\mI-p\sum_{\alpha}L_{\alpha}^{\dagger}L_{\alpha}}{1-p}\right)^{\frac{1}{2}}
\end{equation}
and $p=\lambda \dt$ $(0<p<<1)$ in order to satisfy the CPTP conditions of quantum channels.
In the Lindblad limit, we observe that:
\begin{gather}
\rho^{out} =
(\mI - i\dt H + \lambda \dt K)\rho^{in}(\mI \lambda + \dt K + i\dt H) \nonumber \\ + \lambda \dt \sum_{\alpha}L_{\alpha}\rho^{in} L_{\alpha}^{\dagger}
\end{gather}

from which one obtains:
\begin{equation}
\begin{aligned}    
    \mE_{ijkl} = \delta_{ik}\delta_{lj} - i\dt H_{ik}\delta_{lj} + i\dt \delta_{ik}H_{lj} + \lambda \dt (K_{ik}\delta_{lj} \\ +  \delta_{ik}K_{lj}) 
    + \lambda \dt\sum_{\alpha}(L^{\alpha})_{ik}(L^{\alpha})^{*}_{jl}
\end{aligned}
\end{equation}

For a system with Kraus rank $N+1<<d^2$, the eigendecomposition of ${\mathbf{C}_\mE}_{(i,k),(j,l)}$, close to the Lindblad limit,  will lead to the following eigenspectrum: one relatively large eigenvalue of order 1 whose eigenvector corresponds to the Kraus operator for Hamiltonian evolution and the overall effect of dissipation, some intermediate non-zero eigenvalues of the order $\dt$ whose eigenvectors correspond to the Lindblad operators and rest of the eigenvalues will be zero representing the kernel. Since the Kraus operator containing the Hamiltonian evolution is given by $(\mI - i\dt H + \lambda \dt K)$ one can obtain an estimate of the Hamiltonian $H$ by looking at the antisymmetric part of the eigenvector corresponding to the top eigenvalue of $\mE_{(i,k),(j,l)}$. The Lindblad operators $L_{\alpha}$ can be identified from the eigenvectors corresponding to the non-zero intermediate eigenvalues. 

One can obtain an estimate of $K=-\frac{1}{2}\sum_{\alpha}L_{\alpha}^{\dagger}L_{\alpha}$. Having obtained estimates of $H$, $L_{\alpha}$, and $K$ one can then write down the GKSL generator $G_{ijkl}$. We can further simulate (see Sec. (\ref{sec:Numerical_exp}), and compare the estimated generators with the actual generator. The generator is given by:


\begin{gather}
    G_{ijkl} = \lim_{\dt \to 0} \frac{\mE_{ijkl} - \delta_{ik}\delta_{jl}}{\dt} \nonumber \\
    = (- i H_{ik}\delta_{jl} + i \delta_{ik}H_{jl}) + \lambda (K_{ik}\delta_{jl} +  \delta_{ik}K_{jl})
    \\ \nonumber + \lambda \sum_{\alpha}(L^{\alpha})_{ik}(L^{\alpha})^{*}_{jl}
\end{gather}

Additionally, the behavioral changes in the structure of the eigenspace, when we have our least square estimated channel $\hat \mE$ instead of the true low-rank channel $\mE$ are discussed further.  However, if we form the equivalent of the estimated Choi matrix, with a large enough sample, we will find three different classes of eigenvalues, the largest one closest to 1, $N$ intermediate eigenvalues, and $d^2-N-1$ non-zero eigenvalues. The last group is a finite sample effect, replacing the zero eigenvalues of the ideal channel. Ideally, this group needs to be well below the intermediate group of eigenvalues.

\section{Estimation noise using random matrix theory}


In order to understand the effect of noise on the eigenspectrum of the estimated channel $\hat{\Phi}_{ik,jl}$ we bring in some tools from Random Matrix Theory.
We can write the estimate of our channel as follows:
\begin{gather} \label{eqn:RMT}
    \hat{\Phi}_{ik,jl} = \Phi_{ik,jl} + X_{ik,jl} \\
    \hat{\Phi}_{I,J} = \Phi_{I,J} + X_{I,J}
\end{gather}
Note that we use the notation $\Phi$ to denote a quantum channel to help differentiate between the index notation $(ij,kl)$ and $(ik,jl)$. When using the input-output index notation i.e. $(ij,kl)$ we choose $\mE$ to represent a quantum channel, and with $(ik,jl)$, we alternatively use $\Phi$.
In the Eqn.\eqref{eqn:RMT},  $X_{I,J}$ denotes the noise in the estimation process. We focus on qudit systems of dimension $d$ and suppose that the rank of our channel is $k$. Thus, the actual $\Phi_{I,J}$ has $d^{2} - k$ zero eigenvalues in its spectrum. We consider the projection of $X_{I,J}$ into this $D = d^2 - k$ dimensional kernel subspace - due to noise in estimation the estimated eigenvalues will not be exactly zero but will be distributed about the zero eigenvalue with a certain characteristic width. We think of $X_{I,J}$ as a random matrix with each of the $D^{2}$ elements $X_{I,J} \sim N(0,\frac{a^2}{n})$ where $a^2$ is a pre-factor for the variance and $n$ denotes the total number of samples. Let the normalized frequency distribution of the eigenvalues of $X_{I,J}$ be denoted by $\rho(\lambda)$ with the variance being $\sigma_{\lambda}^{2}$. We observe the following:
\begin{gather}
    \mathbb{E}[\sum_{\alpha=1}^{D}\lambda_{\alpha}^{2}] = D \sigma_{\lambda}^{2} \\
    \mathbb{E}[\sum_{\alpha=1}^{D}\lambda_{\alpha}^{2}] = \mathbb{E}[\text{Tr}(X^2)] = \mathbb{E}[\sum_{I}\sum_{J} X_{I,J}X_{J,I}] = D^{2}\frac{a^{2}}{n} \\
    D \sigma_{\lambda}^{2} = D^{2}\frac{a^{2}}{n} \implies 
    \sigma_{\lambda}^{2} = \frac{D}{n}a^2
\end{gather}

This indicates that the estimated eigenvalues corresponding to the kernel subspace will be distributed about the actual zero eigenvalues with a characteristic width of the order $\sigma_{\lambda} = \sqrt{\frac{D}{n}}a$ and hence these eigenvalues can mix with the intermediate eigenvalues of order $\dt$ corresponding to the Lindblad operators. Thus, one needs a certain optimal number of samples $n$ to actually differentiate between the estimates of the trivial eigenvalues and the non-trivial ones. One important thing to note is that the intermediate eigenvalues will be of the order $||L_{\alpha}||_{F}^{2} \dt$ if $L_{\alpha}$ is not normalized. In order to get the normalized Frobenius norm for the Lindblad operators, one can sample the elements of the matrices $L_\alpha$ from Gaussian distribution and then divide by a factor of $\sqrt d$ so that the Frobenius norm is 1 in expectation. In that case, the gap between the zero eigenvalues and the intermediate ones will be of the order $\dt$. To differentiate between the estimates of trivial and nontrivial eigenvalues in the spectrum one needs to ensure $\sigma_{\lambda} = \sqrt{\frac{D}{n}}a  << \dt$ ie. $n >> \frac{D a^2}{\dt^2}$.

\begin{figure*}[ht]
    \centering
\includegraphics[width=17cm]{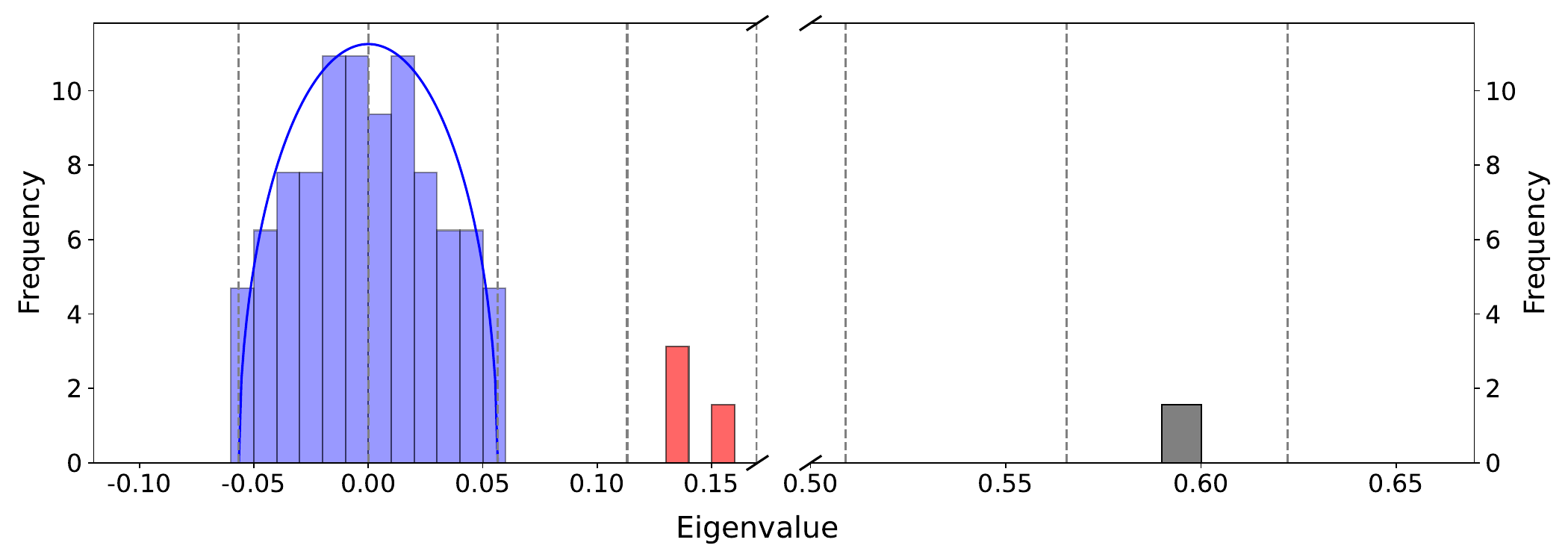}
    \caption{The histogram of eigenvalues for a system of 3-qubits acted upon by a channel of rank 4. With a high probability, the system is evolved according to a Hamiltonian and with the remaining, it is acted by a Pauli channel Eq. \eqref{"simple_H_pauli"}. Here we show that with $10000$ shadows, we can recover the true rank of the channel from the shadow-based estimates.
    The highest eigenvalue is marked in green, intermediate eigenvalues are marked with blue and the noise-induced eigenvalues around zero are marked with orange.
    We also explain the distribution of the low-lying eigenvalues using the Wigner semicircle law (marked in red). The simulation has been run on the IBM-Q QASM simulator.}
    \label{fig:hist_split}
\end{figure*}

\section{Numerical experiments}
\label{sec:Numerical_exp}
\textit{Measurement procedure}:
We use ideas from state tomography to obtain an estimate of $\mE_{ijkl}$. We prepare multiple copies of a set of informationally complete quantum states for qudit-like systems and aim to estimate the output state after the action of a channel by using tools from shadow tomography. As input states we use $d^2$ pure qudit states each of dimension $d$ such that the projectors of these input states form an informationally complete basis. After evolving each state through the channel we use shadow tomography protocol to estimate the output state - we apply a Haar random unitary conjugation on the output state, then measure using the projectors of the computational basis for $\mathbb{C}^d$ and form a shadow like estimate by averaging over many measurement outcomes for each input state. The motivation behind following this protocol is that the inverse of the measurement channel is analytically tractable using the averaging properties of Haar random unitaries. A practical way to implement this is using SIC POVMs, MUBs or Clifford circuits which have the 2-design property. Another measurement protocol is choosing a set of informationally complete POVMs and building shadow estimates based on the measurement outcomes although in this case the measurement channel inverse might not be analytically tractable and has to be implemented numerically.
Since the regression estimate of $\mE$ from the loss function discussed previously is given by $\mE = \mathbf{C}^{out,in}(\mathbf{C}^{in,in})^{-1} $ we see that the estimation error in $\mE$ appears due to noise in the estimation of $\mathbf{C}^{out,in}$. As long as we use a set of informationally complete basis states as our input, $(\mathbf{C}^{in,in})^{-1}$ exists and is already known.
Thus, the randomness/noise in our estimation procedure is manifested through the error in the estimation of the output quantum state.

\begin{figure}
    \centering
    \includegraphics[width=8cm]{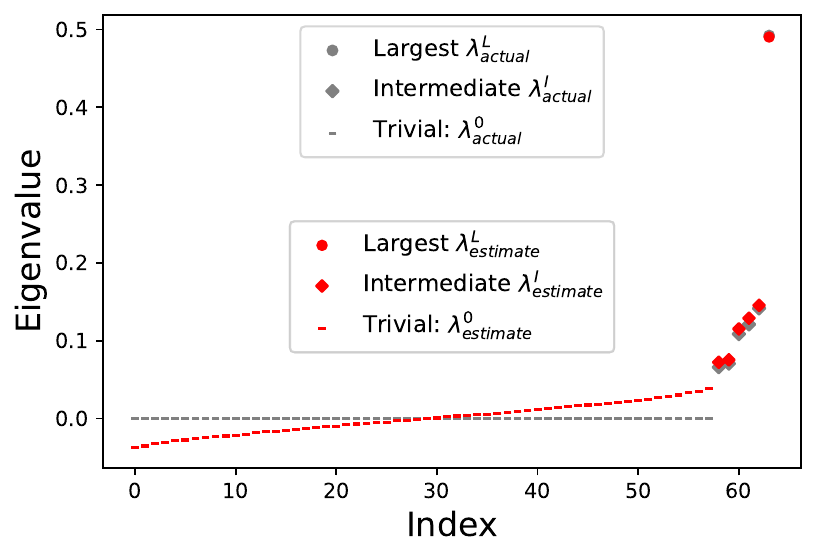}
    \caption{Gaps between the highest, intermediate non-trivial eigenvalues and the noise-induced eigenvalues around zero. The system considered is a qudit of dimension $d=8$ and a channel of rank 4 composed of Hamiltonian evolution and 5 random Lindblad operators. The evolution is for $p=dt=0.1$ and we have chosen $7000$ sample size, so the total number of samples is $64 \times 7000.$ }
    \label{fig:energygap}
\end{figure}
\textit{Experiments}:
As one of the simple examples we consider a system of 3 qubits acted upon by the following channel:
\begin{equation}
    \label{"simple_H_pauli"}
    \mE(\rho) = (1-p)U\rho U^{\dagger} + \sum_{\alpha=1}^{3}\frac{p}{3}L^{\alpha}\rho L^{\alpha \dagger}
\end{equation}

We choose $U$ to correspond to evolution by Hamiltonian of the form $\sigma_{z}\otimes I \otimes I$ and $L_{i}$ denotes $\sigma_{x}$ acting only on the $i$th qubit. It is easy to see that $U$ and $L_{\alpha}$ are unitaries that are orthogonal under the trace inner product. Thus this is a channel of rank 4. Since $p$ is very small we expect one large eigenvalue, three intermediate eigenvalues, and rest of the eigenvalues to be close to zero in the estimated eigenspectrum. We demonstrate the results as a histogram as seen in Fig. \ref{fig:hist_split}.

\begin{figure}
    \centering

\subfloat[ Error in Choi matrix]{\includegraphics[width=8cm]{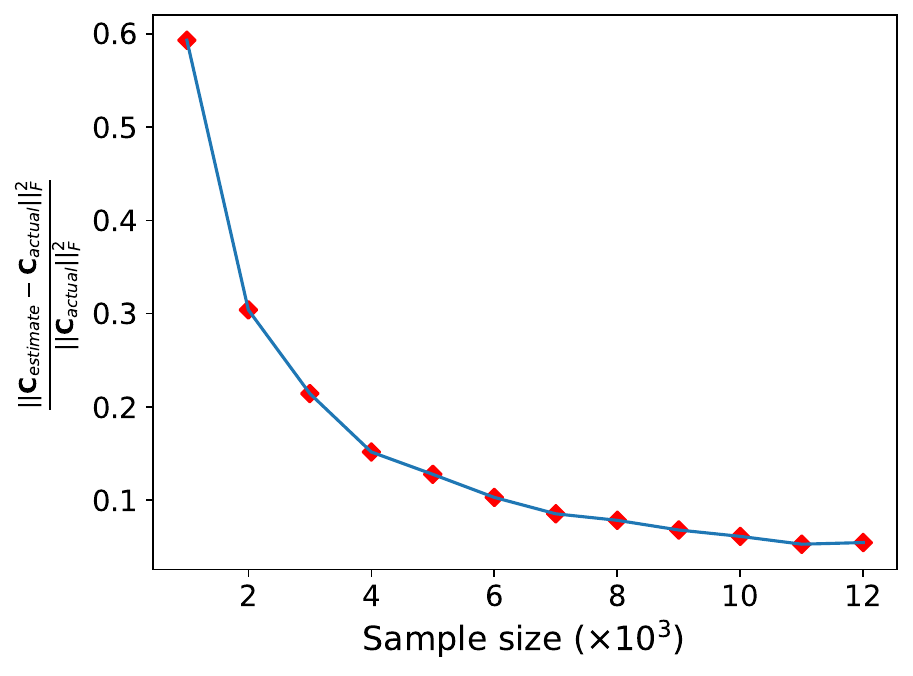} }

\subfloat[ Error in Generators]{\includegraphics[width=8cm]{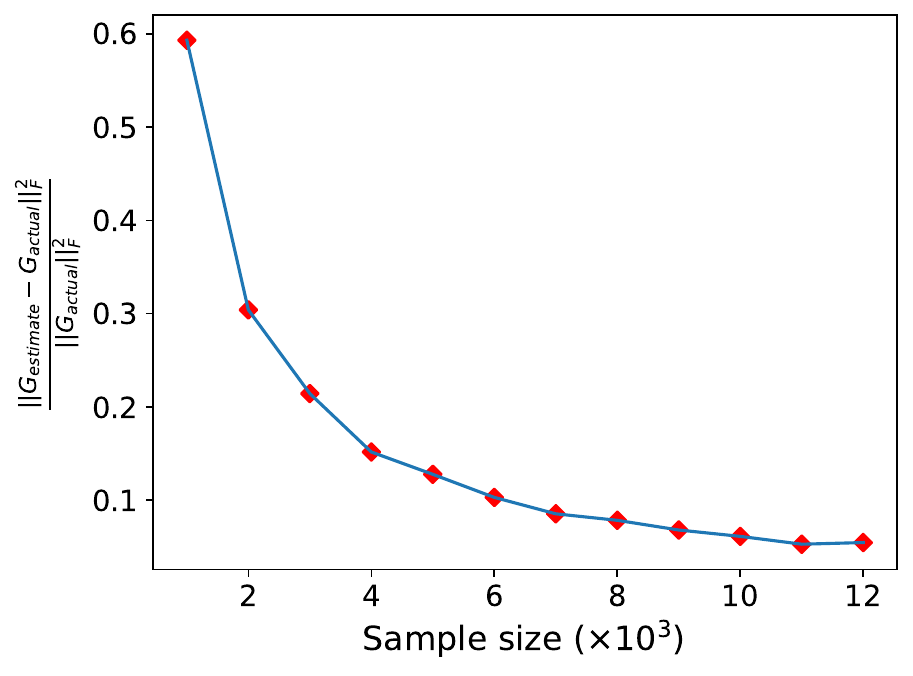} }

\caption{ (a) Plot of the scaled Frobenius error ($\frac{||C_{estimate}-C_{actual}||_{F}^{2}}{||C_{actual}||_{F}^{2}})$ in the estimation of the processed Choi matrix (C) versus sample size. The system considered is a qudit of dimension $d=8$ and a channel of rank 6 composed of Hamiltonian evolution and 5 random Lindblad operators. The evolution is for $p=dt=0.1$. (b) Plot of the scaled Frobenius  error between the true generator $(G_{actual})$ and the estimated generator $(G_{estimate})$ in estimation of the corresponding Generetors}
    
    \label{fig:Choi_gen}
\end{figure}

For our experiments, we consider a qudit of dimension $d = 8$ and a channel
of rank 6 composed of Hamiltonian evolution and 5 Lindblad operators. The Hamiltonian and Lindblad operators are generated randomly. In fig.~\ref{fig:energygap} we show plots of the eigenvalues obtained from the eigendecomposition of the original Choi matrix and the estimated Choi matrix. In fig.~\ref{fig:Choi_gen} we show how the error in estimation of the processed Choi matrix behaves as we increase the sample size. The size of the sampling noise induced eigenvalues are expected to be $O\big(\sqrt{\tfrac{d^2-N}{T}}\big)$, explaining the improvement of reconstruction error with  growing sample size.


Principal component analysis \cite{hotelling1933analysis}, in particular, and truncated singular value decomposition (TSVD) based denoising \cite{golub1965calculating}, in general, are widely-used methods that have led to many theoretical discussions about recovery of low rank-signal from noisy matrices. Tools from the theory of random Wishart matrices played a major role in this discussion \cite{marvcenko1967distribution,sengupta1999distributions,baik2005phase,gavish2014optimal}.  Similar sophisticated tools from random matrix theory could be developed for this problem to discover Lindblad operators, while dealing with sampling noise.

\begin{figure}
    \centering
    \includegraphics[width=8cm]{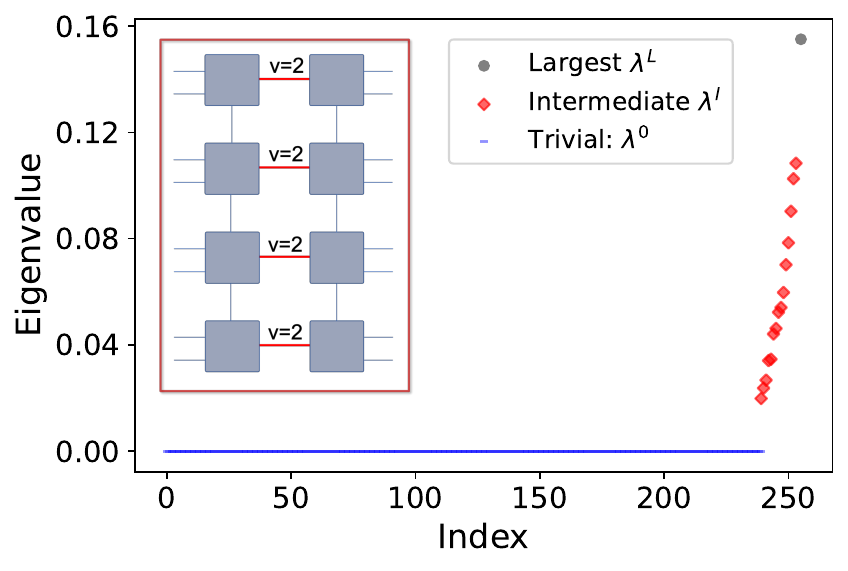}
    \caption{We consider a Choi matrix parametrized by a locally-purified density operator (LPDO) consisting of 4 qubits and a Kruas dimension ($\nu=2$). Thus we observe $2^4-1=15$ intermediate eigenvalues, 1 eigenvalue that is needed for trace preservation and the remaining $240$ trivial eigenvalues.  We compute the spectrum of the Choi matrix obtained after variationally learning the quantum process from measurement data on a random circuit ansatz of depth=10 on a 4 qubit system. This parametrization enables us to scale up process tomography tasks as discussed in \cite{torlai2020quantum}.   }
    \label{fig:LPDO}
\end{figure}

\section{Appendix}
\subsection{The Gorini-Kossakowski-Sudarshan-Lindblad (GKSL) equation}
\label{sec:Appendix_GKSL}

The Kraus operator summation can be used to write the evolution of $\rho$ from $t$ to $t+\delta t$ as: $\rho(t+\delta t)=\sum_{k} M_{k}(\delta t) \rho(t) M_{k}^{\dagger}(\delta t)$.
If we work in the limit of infinitesimal time, $\delta t \rightarrow 0$. The first order survive in $\delta t, \rho(t+\delta t)=\rho(t)+\delta t \delta \rho$. This implies that the Kraus operator should be expanded as $M_{k}=M_{k}^{(0)}+\sqrt{\delta t} M_{k}^{(1)}+\delta t M_{k}^{(2)}+\ldots$ Then there is one Kraus operator such that $M_{0}=\mI+\delta t(-i \mathcal{H}+K)+O\left(\delta t^{2}\right)$ with $K$ hermitian (so that $\rho(t+\delta t)$ is hermitian), while all others have the form: $M_{k}=\sqrt{\delta t} L_{k}+O(\delta t)$, so that we ensure $\rho(t+\delta t)=\rho(t)+\delta \rho \delta t$ :

\begin{gather}
\rho(t+\delta t)=M_{0} \rho(t) M_{0}^{\dagger}+\sum_{k>0} M_{k} \rho M_{k}^{\dagger} \\
=[\mI+\delta t(-i \mathcal{H}+K)] \rho[\mI+\delta t(i \mathcal{H}+K)]+\delta t \sum_{k} L_{k} \rho L_{k}^{\dagger} \\
=\rho-i \delta t[\mathcal{H}, \rho]+\delta t(K \rho+\rho K)+\delta t \sum_{k} L_{k} \rho L_{k}^{\dagger}
\end{gather}

where operator $K$ and the other operators $L_{k}$ are related to each other since they have to respect the Kraus sum normalization condition,
\begin{equation}
K=-\frac{1}{2} \sum_{k>0} L_{k}^{\dagger} L_{k}
\end{equation}
Finally we substitute $K$ in the equation above and take the limit $\delta \rightarrow 0: \rho(t+d t)=\rho(t)+d t \dot{\rho}$. We thus obtain the Lindblad master equation with $\{ A, B\}=AB+BA$
\begin{gather}
\label{Eq:master_eqn1}
\dot{\rho}(t)=-i[\mathcal{H}, \rho(t)]+\sum_{k=1}^{N}\left(L_{k} \rho(t) L_{k}^{\dagger} -\frac{1}{2} \{ L_{k}^{\dagger} L_{k}, \rho(t) \} \right)
\end{gather}

\def\dt{\Delta t}

\subsection{Near unitary channels}
Starting with pure unitary evolution, we can write the quantum channels for small-time evolution as 

\begin{gather}
    U=e^{-i\dt H}
    \approx \mI - i\dt H \\
    U_{ik}=\delta_{ik}-i\dt H_{ik} \quad \text{(matrix \; notation)} \\
    U^{*}_{jl}=\delta_{jl}+i\dt H_{jl}
\end{gather}
We have used $U^{\dagger}_{lj}=U^{*}_{jl}$ and the Hermitian property of Hamiltonian $H^{\dagger}_{jl}=H_{jl}$. This gives us the evolution of the input density matrix term for only a unitary channel as 
\begin{gather}
\rho^{out}=\mE(\rho^{in})=U\rho^{in}U^{\dagger} \\
\implies \rho^{out}_{ij}=\sum_{kl}U_{ik}\rho^{in}_{kl}U^{\dagger}_{lj}=\sum_{kl}U_{ik}U^{*}_{jl}\rho^{in}_{kl}\\
    =(\delta_{ik}-i\dt H_{ik})(\delta_{jl}+i\dt H_{jl})\; \rho^{in}_{kl} \\
    =(\delta_{ik}\delta_{jl}+i\dt H_{jl}\delta_{ik}-i\dt H_{ik}\delta_{jl}+(\dt)^{2}H_{ik}H_{jl} ) \; \rho^{in}_{kl}
\end{gather}

Thus for small-time evolution using the unitary channel, we obtain 
\begin{equation*}
    \mE_{ijkl} = (\delta_{ik}\delta_{jl}+i\dt H_{jl}\delta_{ik}-i\dt H_{ik}\delta_{jl}+(\dt)^{2}H_{ik}H_{jl} )
\end{equation*}
For the generator of the channel, we obtain:
\begin{equation}
    G_{ijkl} = \lim_{\dt \to 0} \frac{\mE_{ijkl} - \delta_{ik}\delta_{jl}}{\dt}  = -iH_{ik}\delta_{jl} + iH_{jl}\delta_{ik}
\end{equation}
Moving from perfect unitaries to near unitaries by adding extra terms. This analysis can be termed as evolution with \textit{mixed-unitary channel}. Simply understood as a convex combination of unitary channels. Note that at the very least they are unital i.e. $\mE(\mI)=\mI$. \cite{watrous_2018}
\begin{gather}
(1-p)U\rho U^{\dagger} + \sum_{\alpha}^{N}p_{\alpha}(L^{\alpha}U\rho U^{\dagger}(L^{\alpha})^{\dagger} )
\end{gather}

The trace condition is satisfied by ensuring $1-p+\sum_{\alpha}p_{\alpha}=1$. For example, depolarizing channel will correspond to uniform distribution i.e. $p_{\alpha}=\frac{p}{N}$.
\begin{gather}
(1-p)(\delta_{ik}\delta_{jl}+i\dt H_{jl}\delta_{ik}-i\dt H_{ik}\delta_{jl} \\ +(\dt)^{2}H_{ik}H_{jl} ) \; \rho^{in}_{kl}+\sum_{\alpha}p_{\alpha}(L^{\alpha}U)_{ik}(L^{\alpha}U)^{*}_{jl}\rho_{kl}   
\end{gather}
If $p=\lambda\dt$, $p_{\alpha}=\lambda_{\alpha}\dt$  and keeping only the first order terms, we get 
\begin{gather}
(1-\lambda\dt)(\delta_{ik}\delta_{jl}+i\dt H_{jl}\delta_{ik}-i\dt H_{ik}\delta_{jl}) \rho^{in}_{kl} \\ +\dt\sum_{\alpha}\lambda_{\alpha}[(L^{\alpha})_{ik}(L^{\alpha})^{*}_{jl} + \mathcal{O}(\dt) ] \; \rho_{kl}\\
\approx [(1-\lambda\dt)\delta_{ik}\delta_{jl}+ (-i\dt H_{jl}\delta_{ik}+i\dt H_{ik}\delta_{jl}) \\ +\dt\sum_{\alpha}\lambda_{\alpha}(L^{\alpha})_{ik}(L^{\alpha})^{*}_{jl}]\;  \rho_{kl}
\end{gather}
We can thus define the generators as; \begin{gather}
    G_{ijkl}= \lim_{\dt \to 0} \frac{\mE_{ijkl} - \delta_{ik}\delta_{jl}}{\dt} \\
    = -\lambda \delta_{ik}\delta_{jl}+ (-i H_{jl}\delta_{ik}+i H_{ik}\delta_{jl})+ \sum_{\alpha}\lambda_{\alpha}(L^{\alpha})_{ik}(L^{\alpha})^{*}_{jl}
\end{gather}
There is additional error coming due to ignoring the higher order terms i.e. $\mathcal{O}((\dt)^2)$

Now let's consider a channel of the following form:
\begin{equation}
    \mE (\rho) = (1-p)UM\rho M^{\dagger}U^{\dagger} + p\sum_{\alpha}L_{\alpha}\rho L_{\alpha}^{\dagger}
\end{equation}

where $L_{\alpha}$ are arbitrary Lindblad-like operators and $M$ has been introduced to satisfy the trace normalization property of CPTP maps.

In order to satisfy the trace condition, we need:
\begin{gather}
    (1-p)M^{\dagger}M + p\sum_{\alpha}L_{\alpha}^{\dagger}L_{\alpha} = \mI \\
    M^{\dagger}M = \frac{\mI-p\sum_{\alpha}L_{\alpha}^{\dagger}L_{\alpha}}{1-p} \\
    M = \left(\frac{\mI-p\sum_{\alpha}L_{\alpha}^{\dagger}L_{\alpha}}{1-p}\right)^{\frac{1}{2}}
\end{gather}

In order to consider the square root in the above equation, the numerator should be positive semi-definite, thus when implementing this procedure we ensure that we choose $p$ and the Lindblad operators $L_{\alpha}$ such that all the eigenvalues of $\mI-p\sum_{\alpha}L_{\alpha}^{\dagger}L_{\alpha}$ are non-negative.

Thus using this expression for $M$ and looking at the action of the channel by expanding till the first order in $\dt$ we obtain:
\begin{equation}
\rho^{out} = \mE(\rho_{in})
\end{equation}

\begin{gather*} 
    = (\mI - i\dt H)(\mI - \frac{1}{2}p\sum_{\alpha}L_{\alpha}^{\dagger}L_{\alpha})\rho_{in}(\mI - \frac{1}{2}p\sum_{\alpha}L_{\alpha}^{\dagger}L_{\alpha}) \\ \times (\mI + i\dt H)
    + p\sum_{\alpha}L_{\alpha}\rho_{in} L_{\alpha}^{\dagger} \end{gather*}
We obtain $(\mI - i\dt H)(\mI + pK)\rho_{in}(\mI + pK)(\mI + i\dt H) + p\sum_{\alpha}L_{\alpha}\rho_{in} L_{\alpha}^{\dagger} $, and then equate it to    
\begin{gather}
    = (\mI - i\dt H + \lambda \dt K)\rho_{in}(\mI + \lambda \dt K + i\dt H) \\
     + \lambda \dt \sum_{\alpha}L_{\alpha}\rho_{in} L_{\alpha}^{\dagger}.
\end{gather}

In the above derivation we used $(\mI-p\sum_{\alpha}L_{\alpha}^{\dagger}L_{\alpha})^{\frac{1}{2}} \approx \mI- \frac{1}{2}p\sum_{\alpha}L_{\alpha}^{\dagger}L_{\alpha}$ since $p$ is of the order $\dt$ and we set $K = -\frac{1}{2}\sum_{\alpha}L_{\alpha}^{\dagger}L_{\alpha}$

Upto first order in $\dt$ we obtain:

\begin{gather*}
    \rho_{ij}^{out} = [\delta_{ik}\delta_{jl} - i\dt H_{ik}\delta_{jl} + i\dt \delta_{ik}H_{jl} + \lambda \dt (K_{ik}\delta_{jl} \\ +  \delta_{ik}K_{jl}) 
    + \lambda \dt\sum_{\alpha}(L^{\alpha})_{ik}(L^{\alpha})^{*}_{jl}] \rho_{kl}^{in}
\end{gather*}

and the generator is given by:

\begin{gather*}
    G_{ijkl} = \lim_{\dt \to 0} \frac{\mE_{ijkl} - \delta_{ik}\delta_{jl}}{\dt} = (- i H_{ik}\delta_{jl} + i \delta_{ik}H_{jl}) \\ + \lambda (K_{ik}\delta_{jl} +  \delta_{ik}K_{jl})
    + \lambda \sum_{\alpha}(L^{\alpha})_{ik}(L^{\alpha})^{*}_{jl}
\end{gather*}

The above equation for the generator is similar to the expression in the Lindblad master equation; the first term corresponds to the commutator term with Hamiltonian, the second term denotes the anti-commutator term with $K$ and the last term corresponds to the action of the Lindblad operators. To estimate the generator from the measurement results, we note that for the channel:

\begin{gather}
    \rho_{out} = \mE(\rho_{in}) = \sum_{\alpha} M_{\alpha} \rho_{in} M_{\alpha}^{\dagger}
\end{gather}

we have the following:
\begin{equation}
    \mE_{ijkl}=\sum_{\alpha}M^{\alpha}_{ik}M^{\alpha^*}_{jl} 
\end{equation}

\subsection{CP and TP projections}
The projection, with respect to the Frobenius norm, of a matrix $X$ onto the set of matrices representing trace-preserving maps is the solution to the following optimization problem:
\begin{gather}
    \text{Proj}_{TP}[X] = \arg \min_{X'} ||X-X^{'}||_{2}^{2} \\
    s.t. Tr_{s}(X') = \frac{1}{d}\mI_{d} 
\end{gather}

The unique solution to the above optimization problem is given by the following closed form expression:
\begin{equation}
    \text{Proj}_{TP}[X] = X + \frac{1}{d}\mI_{d} \otimes (\frac{1}{d}\mI_{d} - tr_{s}(X))
\end{equation}

The projection of a matrix $X$ onto the set of positive-semidefinite matrices is the solution to the following optimization problem:

\begin{gather}
    \text{Proj}_{CP}[X] = \arg \min_{X'} ||X-X^{'}||_{2}^{2} \\
    s.t. X'\succcurlyeq 0
\end{gather}
The condition of positive semidefiniteness is that all eigenvalues be greater than equal to zero. An obvious method, therefore, for enforcing the positive semidefiniteness of a matrix is to set all negative eigenvalues to zero. This turns out to be the unique solution to the above optimization problem. 
\newline

\subsection{Matrix perturbation and Davis Kahan Theorem}
We bring in ideas from the matrix perturbation theory in order to obtain certain probabilistic bounds. As usual, the estimated channel is given by:
\begin{gather}
    \hat{\mE}_{ik,jl} = \mE_{ik,jl} + X_{ik,jl} \\
    \hat{\mE}_{I,J} = \Phi_{I,J} + X_{I,J}
\end{gather}

The noise term $X_{I,J}$ can be viewed as a perturbation to the original matrix. Let us consider the eigendecomposition of the original channel to be:
\begin{equation}
    \Phi_{I,J} = E_{0}D_{0}E_{0}^{\dagger} + E_{1}D_{1}E_{1}^{\dagger}
\end{equation} 

and the eigendecomposition of the estimated channel to be:
\begin{equation}
    \hat{\Phi}_{I,J} = \tilde{E_{0}}\tilde{D_{0}}\tilde{E_{0}}^{\dagger} + \tilde{E_{1}}\tilde{D_{1}}\tilde{E_{1}}^{\dagger}
\end{equation}

Here $E_{0}$ consists of an orthonormal basis that spans the eigenspace corresponding to $D_{0}$ and $E_{1}$ spans the orthogonal complement. Thus $E_{0}E_{0}^{\dagger} + E_{1}E_{1}^{\dagger} = \mI$ and similarly $\tilde{E_{0}}\tilde{E_{0}}^{\dagger} + \tilde{E_{1}}\tilde{E_{1}}^{\dagger} = \mI$. We would like to figure out how closely the subspace spanned by $\tilde{E_{0}}$ approximates the subspace spanned by $E_{0}$ and thus minimize $||\tilde{E_{0}}\tilde{E_{0}}^{\dagger} - E_{0}E_{0}^{\dagger}||_{F}^{2}$ which is equivalent to minimizing $||\tilde{E_{1}}^{\dagger}E_{0}||_{F}^{2}$. Using the Davis-Kahan theorem we obtain:

\begin{equation}
    ||\tilde{E_{1}}^{\dagger}E_{0}||_{F}^{2} \le \frac{||\tilde{E_{1}}^{\dagger} X E_{0}||_{F}^{2}}{g^2}
\end{equation}

Here $g$ denotes a number such that if the eigenvalues corresponding to $E_{0}$ are contained in the interval $[a,b]$, then the eigenvalues corresponding to $\tilde{E_{1}}$ are excluded from the interval $(a-g,b+g)$.
Note that $||\tilde{E_{1}}^{\dagger} X E_{0}||_{F}^{2} \le ||X||_{F}^{2}$. Thus, if we want to increase the bound $||\tilde{E_{1}}^{\dagger}E_{0}||_{F}^{2}$ by $\mE$, it suffices to increase the bound $||X||_{F}^{2}$ by $g^{2}\mE$.
Applying Markov's inequality on the random variable $||X||_{F}^{2}$ we obtain:
\begin{equation}
    \text{Pr}(||X||_{F}^{2} \le g^{2}\mE) > 1 - \frac{E[||X||_{F}^{2}]}{g^{2}\mE}
\end{equation}




\bibliographystyle{plain}

\bibliography{papers}

\end{document}